\documentclass[conference]{IEEEtran}

\usepackage{etoolbox}
\makeatletter
\patchcmd{\@makecaption}
  {\scshape}
  {}
  {}
  {}
\makeatother

\IEEEoverridecommandlockouts
\usepackage{cite}
\usepackage{amsmath,amssymb,amsfonts,gensymb}
\usepackage{algorithmic}
\usepackage{graphicx}
\usepackage{textcomp}
\usepackage{pifont}
\def\BibTeX{{\rm B\kern-.05em{\sc i\kern-.025em b}\kern-.08em
    T\kern-.1667em\lower.7ex\hbox{E}\kern-.125emX}}

\usepackage[table]{xcolor}
\usepackage{makecell}
\usepackage{siunitx}
\usepackage{multirow}

\graphicspath{{./sections/figs/}}

\title{Towards Multidimensional Textural Perception and Classification Through Whisker\\}

\author{Prasanna Kumar Routray${}^1$, Aditya Sanjiv Kanade${}^1$, Pauline Pounds${}^2$, Manivannan Muniyandi${}^1$

\thanks{${}^1$Touch Lab, Center for Virtual Reality and Haptics, Indian Institute of Technology Madras, India}

\thanks{${}^2$School of Information Technology and Electrical Engineering, The University of Queensland, Brisbane, Australia}

\thanks{(Prasanna Kumar Routray and Aditya Sanjiv Kanade contributed equally to this work.) (Corresponding author: Prasanna Kumar Routray; prasanna.routray97@gmail.com)}
}

\begin{document}

\maketitle

\begin{abstract}
Texture-based studies and designs have been in focus
recently. Whisker-based multidimensional surface texture data is
missing in the literature. This data is critical for robotics and
machine perception algorithms in the classification and regression of
textural surfaces. In this study, we present a novel sensor design
to acquire multidimensional texture information. The surface
texture’s roughness and hardness were measured experimentally
using sweeping and dabbing. Three machine learning models
(SVM, RF, and MLP) showed excellent classification accuracy for
the roughness and hardness of surface textures. We show that
the combination of pressure and accelerometer data, collected
from a standard machined specimen using the whisker sensor,
improves classification accuracy. Further, we experimentally validate that
the sensor can classify texture with roughness depths as low as
$2.5\mu m$ at an accuracy of $90\%$ or more and segregate materials
based on their roughness and hardness. We present a novel metric
to consider while designing a whisker sensor to guarantee the
quality of texture data acquisition beforehand. The machine learning model
performance was validated against the data collected from the
laser sensor from the same set of surface textures. As part of our
work, we are releasing two-dimensional texture data: roughness
and hardness to the research community.

\end{abstract}

\section{Introduction}
The sense of touch is crucial in determining surface texture. Rats exhibit a significantly different sense of touch compared to primates who use their fingers predominantly. Whiskers are used extensively for examining, exploring, and navigating through environments by rats \cite{manfredi2014natural}. The spatiotemporal information is captured at the base of the whisker and subsequently encoded in the barrel cortex of the rat brain. Inspired by whisker-dependent touch sensing, researchers are increasingly concentrating on developing novel sensing technologies for textural perception.

\begin{figure}[ht]
    \centering
    \includegraphics[width=0.33\textwidth]{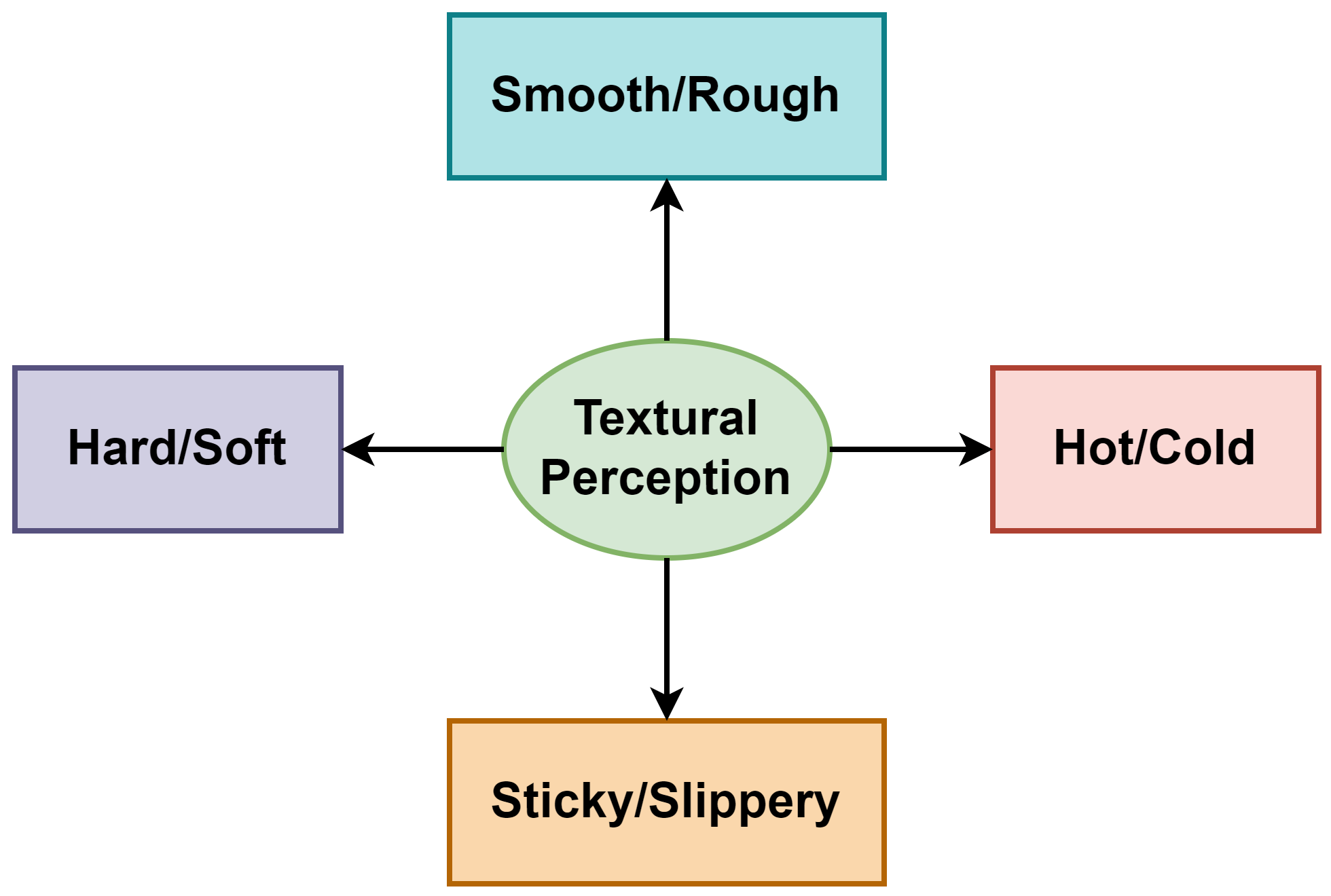}
    \caption{Dimensions of texture, adapted from \cite{okamoto2012psychophysical}}
    \label{fig:textureDimension}
\end{figure}

Rats can distinguish textures as small as $30\mu m$ \cite{birdwell2007biomechanical}.Therefore, sensor designs inspired by rat whiskers are built accordingly with at least $30\mu m$ resolution to discriminate between the rough and slippery surfaces. Studies show that a single whisker can discriminate texture roughness by capturing temporal features \cite{hipp2006texture}. As texture can only be acquired by relative movement between sensor and surface: temporal data, spectrogram of the temporal data, and velocity spectrogram are crucial for decoding the sensor information \cite{arabzadeh2005neuronal}. The relative motion between the sensor and the surface causes micro-movements at the whisker base due to surface coarseness. The slip-stick mechanism is widely accepted as a cause of these micro-movements, which encode the frequency information of the texture \cite{wolfe2008texture}. 
Previous works have focused only on determining texture roughness \cite{jadhav2010texture,hipp2006texture}. However, the literature has not adequately addressed the multi-dimensional nature (see Fig. \ref{fig:textureDimension}) of texture \cite{okamoto2012psychophysical}.


Few studies have designed a whisker sensor for texture classification based on roughness alone \cite{lungarella2002artificial, ju2014bioinspired}. The mean speed of temporal information captured by the whiskers is used for texture classification. Recently, terrain classification has been implemented using a simple IMU accelerometer-based whisker sensor \cite{giguere2009surface}. A bayesian nonparametric approach, similar to the Dirichlet process, is used for texture perception spanning 28 different surfaces \cite{dallaire2014autonomous}. However, the multi-dimensional texture perception is not addressed in these studies, except for the roughness

The main aim of this study is threefold: (1) to design a lightweight whisker sensor for capturing multi-dimensional texture data: roughness and hardness; (2) to evaluate texture roughness and hardness classification using whisker sensor data; and 3) to make the whisker sensor texture data available for the research community. This will help develop test cases for robotics and machine perception algorithms, in addition to studying multivariate time-series data in the domain of classification and regression.

We present a novel design of the whisker sensor based on a pressure sensing mechanism with an accelerometer at the base for collecting multi-dimensional texture data. We show that the combination of pressure and accelerometer data results in improved texture roughness classification over a surface with grain size as small as $2.5 \mu m$. We also introduce a simple metric to be considered while designing a whisker sensor that ensures the quality of texture data. The lack of whisker-based surface texture data hinders further research in the field. As a part of our work,  we also release texture data in two dimensions: roughness and hardness.
\section{Experimental Setup}
A simple whisker sensor has been designed that is capable of acquiring multidimensional surface texture data. Inspired by the macrovibrissae and previous design \cite{deer2019lightweight}, we built a new whisker sensor board for terrain texture mapping. The new whisker board houses an MPU9250, which is a $9DOF$ IMU. To further reduce the sensor’s cost and simplify the fabrication process, we have used silicone rubber of shore $20A$ hardness for load pad instead of polyurethane. It exceeds the design requirements specified in the previous work and, with a new micro-controller (STM32F412CGU6), can accommodate more sensors if needed.


\begin{figure}[ht]
\centering
\includegraphics[width=0.46\textwidth]{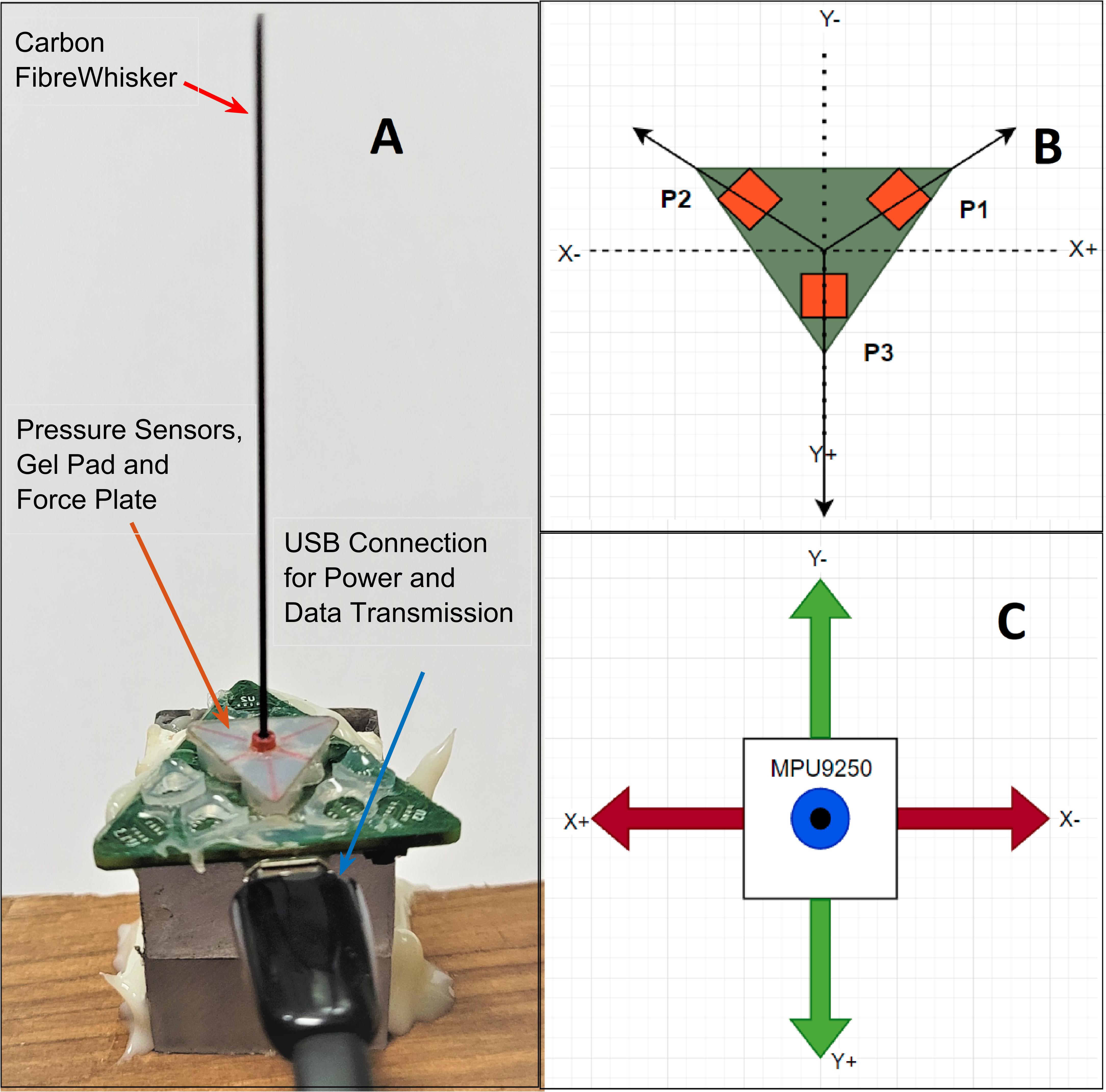}
\caption{Whisker sensor setup and reference frame. (A) The pressure sensor, accelerometer and carbon fibre based whisker sensor, (B) $120\degree$ basis reference frame and Cartesian reference frame for the whisker sensor, (C) Cartesian reference frame for the accelerometer embedded inside MPU9250. Z-Axis is perpendicular to the plane of the paper}
\label{fig:WhiskerSetup}
\end{figure}

The whisker sensor fabricated can be seen in Fig.\ref{fig:WhiskerSetup}(A). The coordinate frames of pressure sensors in cartesian space and the accelerometer are given in Fig.2\ref{fig:WhiskerSetup}(B) \& (C), respectively. A custom XYZ linear stage with $1\mu m$ resolution is used for experimental data collection. An Optical NCDT laser sensor by micro-epsilon captures textural roughness as a ground truth sampled at $50 kHz$ for . The distal range of Optical NCDT ranges from $24-26 mm$ with a resolution of $0.03\mu m$. This is a class-2 laser with a red beam that operates at $670 nm$.


\subsection{Experimental constraints}
\label{sec:constraints}
Texture being a perceptual quantity, necessitates multimodal data to be captured. This data can then be used for texture classification and regression tasks. In order to acquire multi-dimensional texture information, we discuss a few experimental constraints on our data collection setup.

\begin{figure}[ht]
\centering
\includegraphics[width=0.45\textwidth]{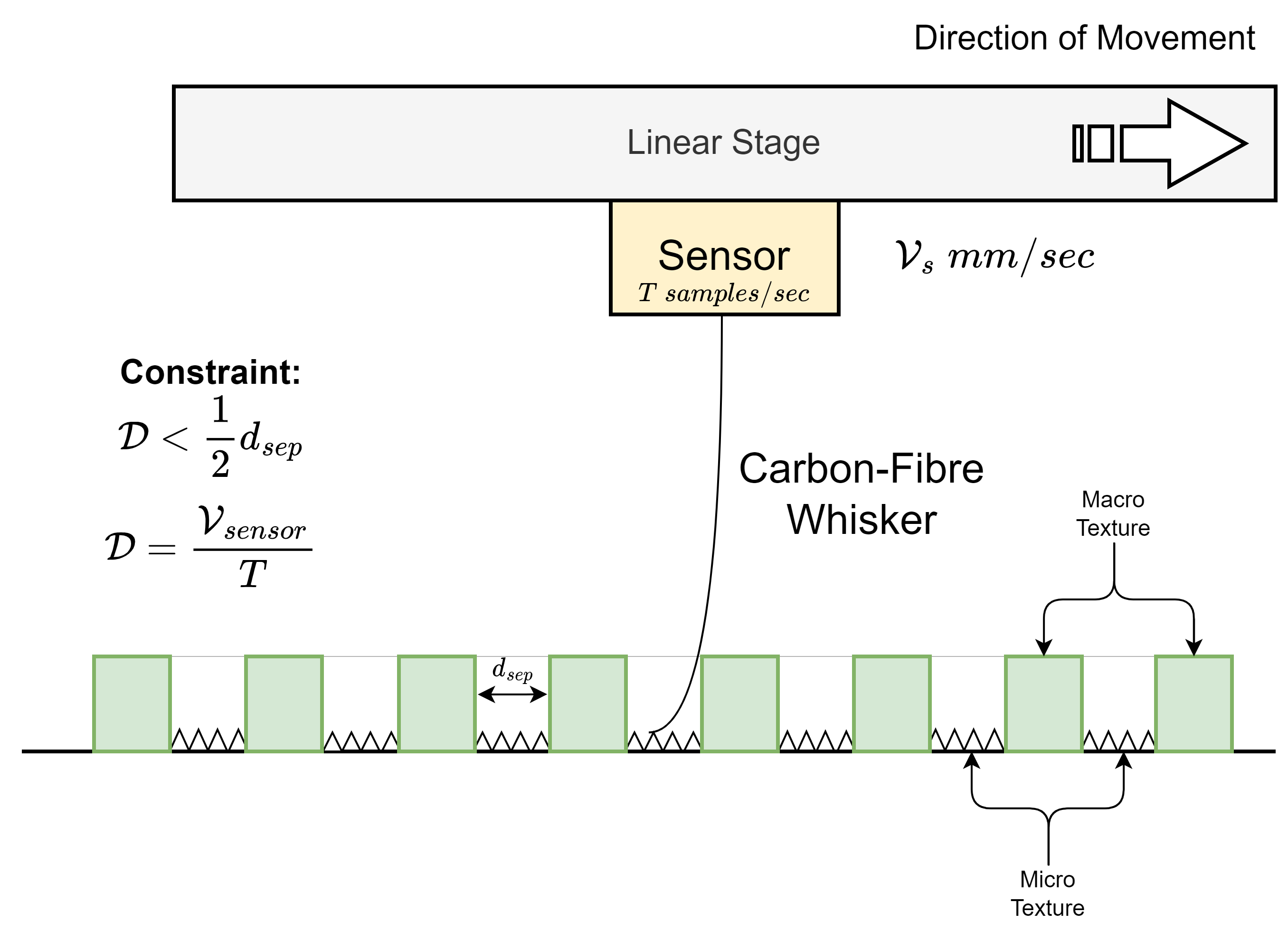}
\caption{A schematic representing the experimental setup and associated constraints. Speed of the linear stage, sensor sampling rate play a critical role in determining minimum measurable texture grain size.}
\label{fig:textureGrainSize}
\end{figure}

Let $N_{s}$, $N_{a}$, and $N_{l}$ denote data rates of the whisker, accelerometer, and laser sensor, respectively. $\mathcal{V}_{s}$ indicates the speed of the linear stage over the texture. As shown in Fig \ref{fig:textureGrainSize}, the minimum distance between two consecutive macro grains of the surface texture is given by $d_{sep}$. A sensor moving in the linear stage covers a distance $\mathcal{D}$ ( refer eq. \ref{eq:D}) between two consecutive samples. Accordingly, an important constraint placed on sensor and the linear stage for capturing high quality texture data is:  $\mathcal{D} < \frac{1}{2}d_{sep}$.





\begin{equation}
\label{eq:D}
\mathcal{D} = \frac{\mathcal{V}_{s}}{\mathcal{N}}   
\end{equation}





Although the laser sensor is capable of capturing surface roughness at higher linear stage speed $\mathcal{V}_{s}$, it is set to match the experimental case of whisker sensor. Special care is taken not to introduce external vibration to avoid noise in range of $\mu m$ measurements.

\subsection{Standard Material selection}
\begin{figure}[ht]
\centering
\includegraphics[width=0.48\textwidth, height=0.4\textheight]{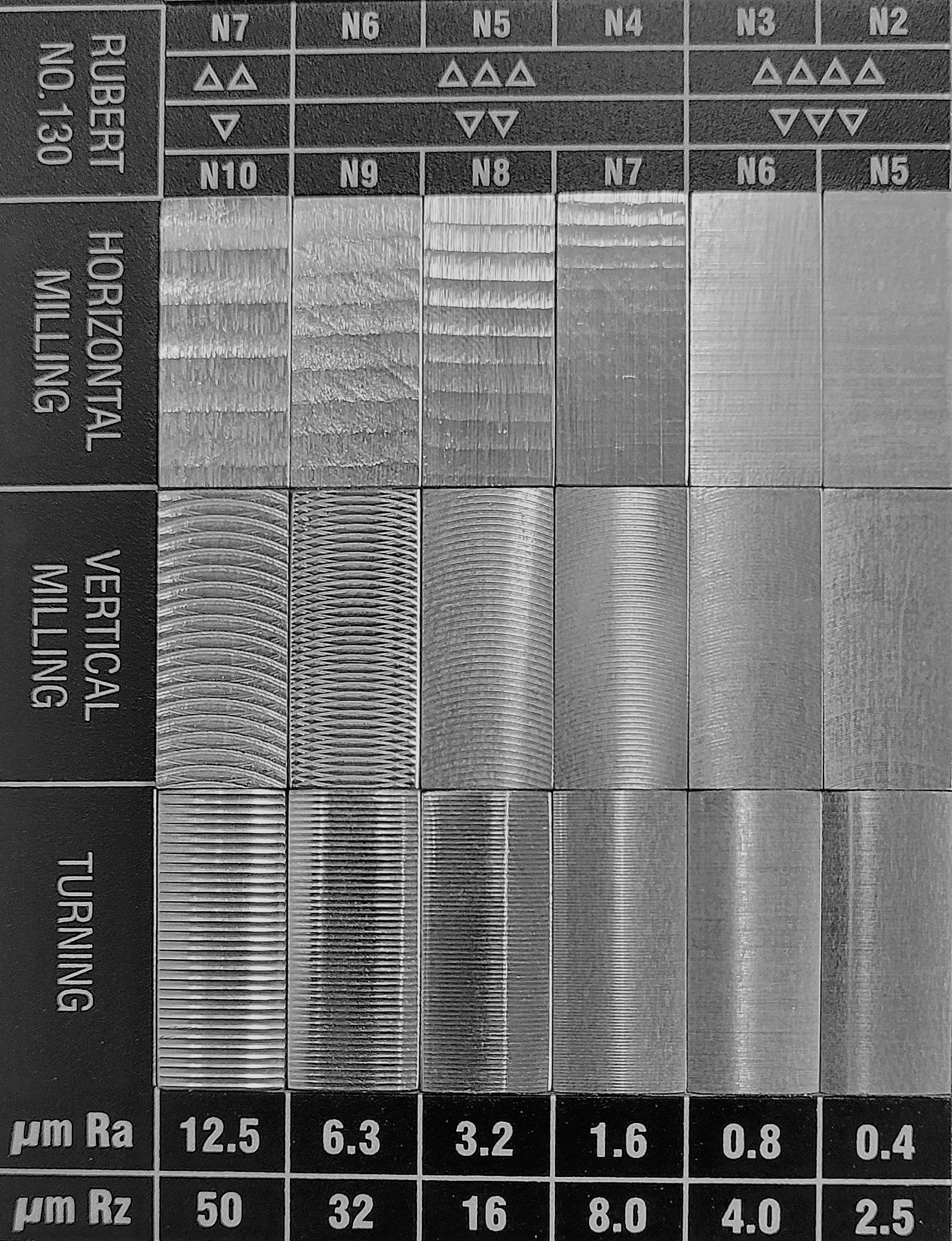}
\caption{A specimen for textural roughness. The peak-to-valley height \textit{Rz} ranges from \SI{2.5}{\mu\meter} to \SI{50}{\mu\meter} with \textit{Ra} values ranging between \SI{12.5}{\mu\meter} to \SI{0.4}{\mu\meter}. It contains 3 classes (vertical milling, horizontal
milling and turning) of steel surface. Each class has six subclasses of roughness.}
\label{fig:roughTexture}
\end{figure}

The materials selected for this task include standard milled and turned metal surfaces with small grain sizes ranging from \SI{2.5}{\mu\meter} to \SI{50}{\mu\meter}, as shown in Fig. \ref{fig:roughTexture}. This machined specimen by RUBERT \& Co. LTD., England, is considered as a standard for surface roughness. The specimen includes three classes of surface roughness: abbreviated with ’H’ for horizontal milling, ’V’ for vertical milling, and ’T’ for turning in this work. Each class has six subclasses of roughness. These eighteen classes of specimens comprise both micro and macro roughness. A grain size above $100\micro m$ is considered a macro, and below is considered a micro.


The average peak-to-valley range \textit{Rz} (Fig. \ref{fig:roughTexture}) is the arithmetic mean of the single roughness depths of successive sample lengths. The mean roughness, \textit{Ra} is  arithmetic mean of the roughness profile ordinates' absolute values. It provides a general description of the surface's height variability. \textit{Ra} is given in expression \ref{eq:roughness} Where L is the length of the specimen and $Z(x)$ is the roughness profile ordinate\cite{chang2001role}.


\begin{equation} \label{eq:roughness}
    \mathcal{R}_{a} = \frac{1}{L}\int_{0}^{L} \lvert Z(x)\rvert \,dx
\end{equation}

\begin{table}[h!]
\centering
\caption{The speed of sensor attached to the linear stage and the sampling rate are key factors in enabling the sensor for obtaining textural roughness information(\ref{eq:D}).}
\scalebox{0.9}{
    \begin{tabular}{|c|c|c|c|c|c|}
    \hline
    \\[-1em]
    \multirow{2}{*}{Sensor Type} & \multirow{2}{*}{$\mathcal{N}$ (Hz)} & \multicolumn{2}{|c|}{$\mathcal{V}_{s}=50\; mm/min$} & \multicolumn{2}{|c|}{$\mathcal{V}_{s}=100\; mm/min$}\\ \cline{3-6}
    \\[-1em]
    & & $\mathcal{D} (\mu m)$ & $d_{sep} (\mu m)$ & $\mathcal{D} (\mu m)$ & $d_{sep} (\mu m)$\\ \hline
    \\[-1em]
    Pressure Sensor & $157$ & $5.31$ & $10.62$ & $10.62$ & $21.23$\\ \hline
    \\[-1em]
    Accelerometer & $1000$ & $0.83$ & $1.67$ & $1.67$ & $3.33$\\ \hline
    \\[-1em]
    NCDT Laser & $2500$ & $0.33$ & $0.67$ & $0.67$ & $1.33$\\ \hline
    \end{tabular}
    }
\label{table:grainSize}
\end{table}

For our experiments, we have chosen a multi-textured surface with a minimum grain size of $2.5\mu m$. From Table \ref{table:grainSize} we found that the pressure sensor and accelerometer combination is enough to capture the detailed texture of the surface under test. We have also collected data using the laser sensor with $N_l = 2500 Hz$ for a comparative reference. 

\begin{figure}[ht]
\centering
\includegraphics[width=0.3\textwidth]{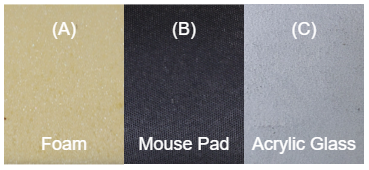}
\caption{A subset of specimens for textural hardness: (A) Soft foam (hardness-1), (B) Table-top mouse pad (hardness-2), (C) Acrylic glass surface (hardness-3)}
\label{fig:materialHardness}
\end{figure}

The other dimension of texture addressed in this study is the hardness of a specimen that can be classified according to the measure of hardness in ascending or descending order \cite{walley2012historical}. We have considered a set of six nonstandard materials (soft foam, a cotton cloth, a mouse pad, double-sided foam tape, cellophane tape, and a painted aluminum sheet) as textural hardness specimens (Fig. \ref{fig:materialHardness}).

\begin{figure}[ht]
\centering
\includegraphics[width=0.46\textwidth]{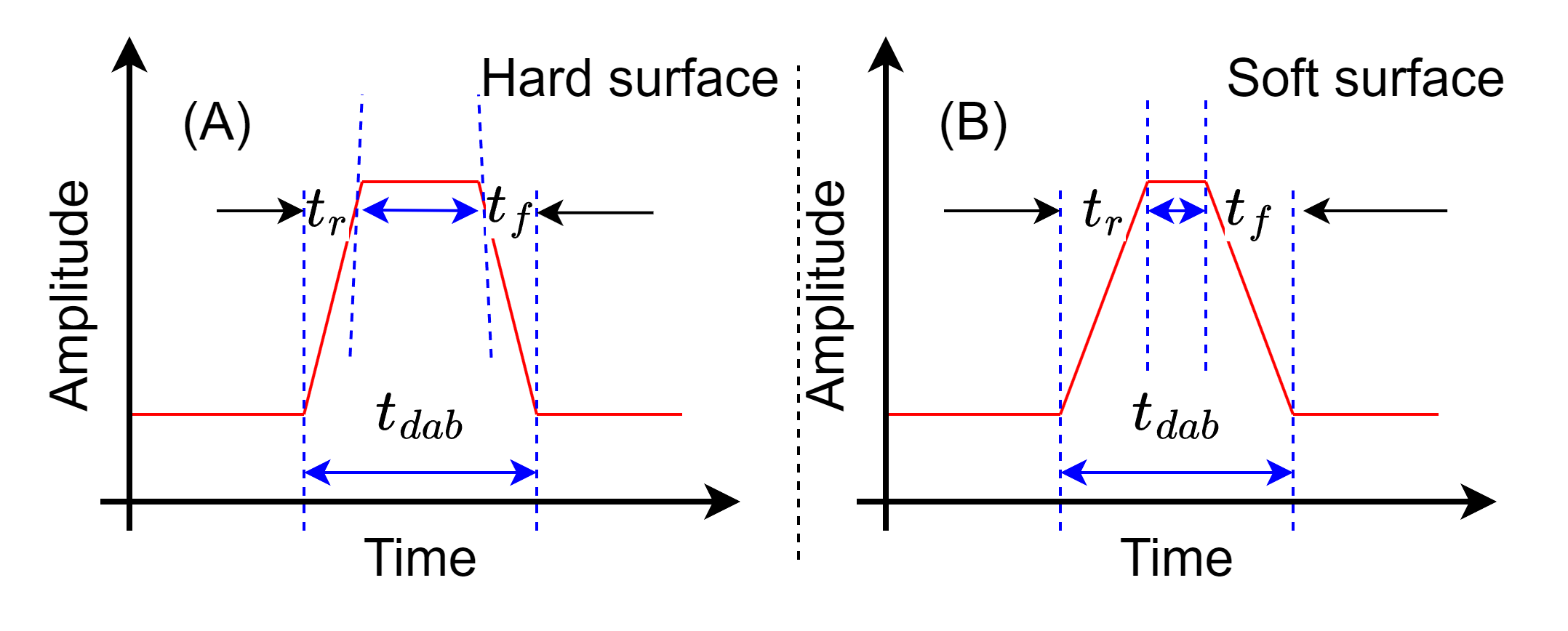}
\caption{Representation of sensor response for: (A) Hard material, (B) Relatively softer material. $t_{dab}$ is the dabbing time starting from whisker making contact to loosing contact with the specimen. For a given fixed dabbing time $t_{dab}$, the rise time $t_{r}$ \& the fall time $t_{f}$ are higher for a relatively softer surface.}
\label{fig:hardnessResponse}
\end{figure}

The slip-stick mechanism responsible for capturing texture roughness is no longer applicable for determining surface hardness. We use the dabbing method to analyze a material’s hardness. Fig. \ref{fig:hardnessResponse} shows the ideal temporal response of the whisker sensor when it makes a vertical dabbing mode of contact with the surface. The response from the sensor during hard surface dabbing reaches the steady state faster than the soft surface. For a hard surface, the rise time is relatively short and higher for a softer surface.



\subsection{Data Collection \& preprocessing}

We collected the data by placing the sensors (whisker and laser) on the linear stage, followed by sweeping and dabbing the specimens for surface texture roughness and hardness, respectively. The constraints described in section II-A are ensured for accurate time-series data. The whisker sensor board streams multivariate data to the PC at 1000 samples/sec. Three sweeps for each class of surface have been collected as a minimum number of three sweeps is required for the classification process \cite{sullivan2011tactile}. The contacts are made by dabbing and sweeping vertically over different class of surfaces spanning multiple texture dimensions. Fig. 3 shows the setup for data collection with a linear moving stage.

\section{Experimental Methodology}
The experimental setup aims to collect texture data with the lightweight whisker and use the data in textural classification on roughness and hardness.

\subsection{Notations}
\label{sec:notations}
\begin{figure}[ht]
\centering
\includegraphics[width=0.48\textwidth]{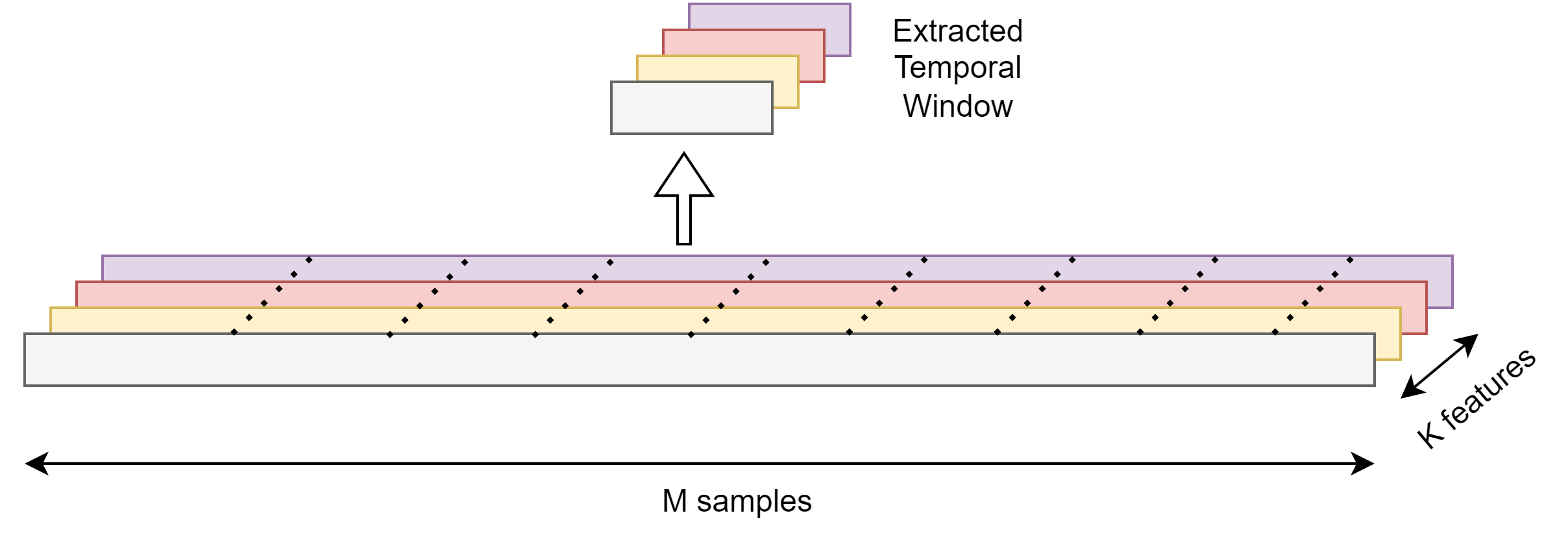}
\caption{Schematic representation of the time-series data and the temporal window blocks that are fed to the machine learning algorithms.}
\label{fig:temporalWindow}
\end{figure}

Let $N \:samples/sec$ be the data rate of the sensor. Let $x \in \mathbf{R}^{k}$ represent the k-dimensional feature collected by the sensor for each sample. The tensor $X = \{x_1, x_2, ..., x_M\}$ represents the time-series data collected over a surface in a sweep. Training machine learning models with the entire sweep data can lead to higher training and inference times hence, we split the tensor $X$ into smaller temporal windows of size $W$, where $X^{j}_{W} = \{x_{j}, x_{j+1}, ..., x_{j+W-1}\}$ as shown in Fig. \ref{fig:temporalWindow}. The window size W is a hyperparameter, and an optimal value is found through experimental analysis (Fig. \ref{fig:acuracyVsInference}).


\subsection{Textural roughness classification}
\label{sec:texture-classification}
Texture data can be used to measure the textural roughness with machine learning models \cite{giguere2009surface, giguere2011simple}. We implement three machine learning models for texture roughness classification. To evaluate the quality of data captured by the new sensor, we also compare the performance of machine learning models trained on data from both whisker and laser sensor.

\subsubsection{Machine Learning Model Setup}
\label{sec:machine-learning-setup}
Three machine-learning methods have been selected for the classification task: 1) Support Vector Machine classifier (SVM), 2) RandomForest classifier (RF), and 3) Multi-Layer Perceptron classifier (MLP). All the models are trained on 0.7/0.2/0.1 train/validation/test split. We have used SVM and RF implementations out of the box from the \textit{scikit-learn} package \cite{scikit-learn}. We used the \textit{TensorFlow} \cite{tensorflow2015-whitepaper} package to implement the MLP model. The MLP has been designed with a stack of dense layers, each with a ReLU activation function; a dropout layer follows each layer to avoid overfitting \cite{dropout}. The input to this network is flattened, followed by a LayerNormalization operation \cite{layer-norm}. The number of perceptrons in the final layer equals the number of classes to predict. This layer has a softmax activation such that it outputs a probability distribution of the model belief on the class of the texture. The MLP-based model is trained using gradient descent with Adam optimizer \cite{adam}. We train the model for 300 epochs with a learning rate set to $0.001$. EarlyStopping \cite{early-stopping} mechanism has been used to avoid overfitting the training data.


\subsection{Textural hardness classification}
Similar to the Texture roughness classification, surface hardness information helps map the environment alongside visual data. Hardness is measured by vertically dabbing the surface. Since there exists no bench-marking device, we categorize the surface based on human perception. For our study, we have considered six specimens of day-to-day life surfaces: acrylic glass, soft cotton, spongy double-side tape, mouse pad, and soft foam.

We used the same machine learning model architecture described in \ref{sec:machine-learning-setup} for textural hardness classification with changes in number of classes for prediction.



\section{Results and discussion}

\begin{table*}[ht]
\centering
\caption{Textural Roughness Classification Results}
\scalebox{0.92}{
\begin{tabular}{|*{18}{c|}}  
\hline
\\[-1em]
\multirow{3}{*}{window size} & \multirow{3}{*}{Classifier} & \multicolumn{8}{|c|}{$\mathcal{V}_{s} = 100$} & \multicolumn{8}{|c|}{$\mathcal{V}_{s} = 50$}\\
\cline{3-18}
\\[-1em]
& &\multicolumn{2}{|c}{\cellcolor{red!10}P}& \multicolumn{2}{|c}{\cellcolor{green!10}A}& \multicolumn{2}{|c}{\cellcolor{blue!10}PA}& \multicolumn{2}{|c}{\cellcolor{yellow!10}L}& \multicolumn{2}{|c}{\cellcolor{red!10}P}& \multicolumn{2}{|c}{\cellcolor{green!10}A}& \multicolumn{2}{|c}{\cellcolor{blue!10}PA}& \multicolumn{2}{|c|}{\cellcolor{yellow!10}L} \\ \cline{3-18}
\\[-1em]
& &$\mu$ &$\sigma^2$ &$\mu$ &$\sigma^2$ &$\mu$ &$\sigma^2$ &$\mu$ &$\sigma^2$ &$\mu$ &$\sigma^2$ &$\mu$ &$\sigma^2$ &$\mu$ &$\sigma^2$ &$\mu$ &$\sigma^2$\\ \hline
\\[-1em]
\multirow{3}{*}{50}& SVM& 77.67& 0.07& 34.85& 0.26& 83.88& 0.14& 36.88& 0.31& 82.51& 0.18& 62.09& 0.16& 92.60& 0.13& 37.36& 0.09\\ \cline{2-18}
\\[-1em]
& RF& 84.64& 0.36& 41.42& 0.08& 86.61& 0.08& 24.46& 0.54& 88.19& 0.02& 60.64& 0.09& 89.25& 0.20& 26.02& 0.07\\ \cline{2-18}
\\[-1em]
& MLP& 46.97& 4.55& 50.61& 0.69& 71.64& 0.30& 87.28 & 0.58 & 54.71& 0.41& 68.18& 1.11& 80.93& 0.53& 90.47& 1.01\\ \cline{1-18}
\\[-1em]
\multirow{3}{*}{100}& SVM& 77.14& 1.43& 28.85& 0.87& 80.93& 1.18& 37.38& 0.34& 82.18& 0.51& 56.61& 0.52& 91.37& 0.05& 37.12& 0.24\\ \cline{2-18}
\\[-1em]
& RF& 84.30& 0.32& 36.01& 2.27& 84.40& 0.57& 23.73& 0.50& 86.87& 0.20& 54.52& 1.31& 87.95& 0.38& 24.64& 0.75\\ \cline{2-18}
\\[-1em]
& MLP& 47.21& 1.51& 47.13& 2.03& 58.41& 1.61& 92.42& 1.64& 55.79& 0.34& 65.86& 0.75& 77.92& 0.99& 92.85& 0.64\\ \cline{1-18}
\end{tabular}
}
\label{tab:resultsRoughness}
\end{table*}

\subsection{Test cases}

\subsubsection{Textural roughness classification}

The designed whisker sensor has a pressure sensor attached to the base of the whisker, which directly records surface texture due to whisker movement. In contrast, the accelerometer, which isn't connected to the whisker, captures an indirect measurement of the surface texture due to induced vibrations resulting from the sweep. Therefore, there is a certain amount of correlation between the data from these two sensors. Based on table \ref{table:grainSize}, it is clear that the pressure sensor is inadequate to capture the minimum grain size of $2.5\mu m$ of the standard specimen. At the same time, the accelerometer has a much higher resolution. We propose combining pressure and accelerometer data to fill this gap in capturing the texture information. We evaluate three machine learning models discussed in section \ref{sec:machine-learning-setup} for the texture classification task. To establish the importance of the combination of both pressure sensor (P) and accelerometer (A) data, we show results in three cases: 1) P, 2) A,  and 3) PA.

\begin{figure}[ht]
    \centering
    \includegraphics[width=0.46\textwidth]{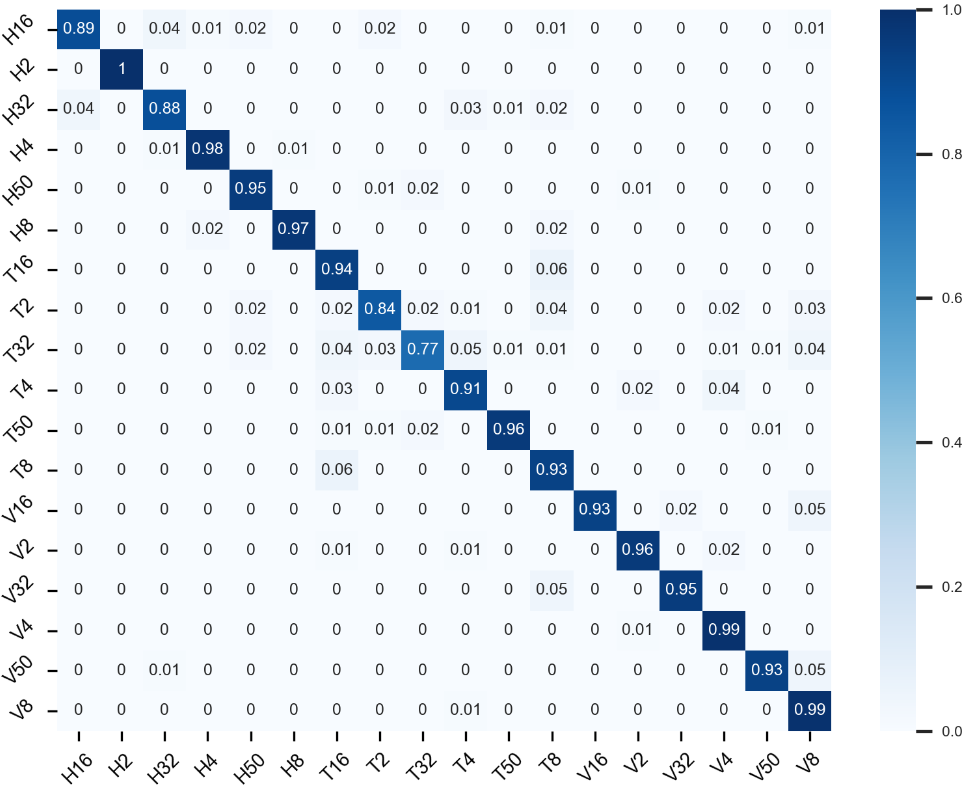}
    \caption{Roughness confusion matrix}
    \label{fig:roughnessConfusionMatrix}
\end{figure}

Table \ref{tab:resultsRoughness} demonstrates that the machine learning models trained on whisker sensor data produces promising texture surface classification results, with excellent classification across all categories as shown in Fig. \ref{fig:roughnessConfusionMatrix}. We see that the PA combination performs the best across all models, indicating that the combination of pressure sensor and accelerometer data yields a much higher textural roughness classification accuracy. It is also crucial to note that as the speed of the linear stage increases ($\mathcal{V}_{s}$), the distance covered by the sensor over the texture between two consecutive samples increases, resulting in a performance decrease across all models.

\begin{figure}[ht]
    \centering
    \includegraphics[width=0.46\textwidth]{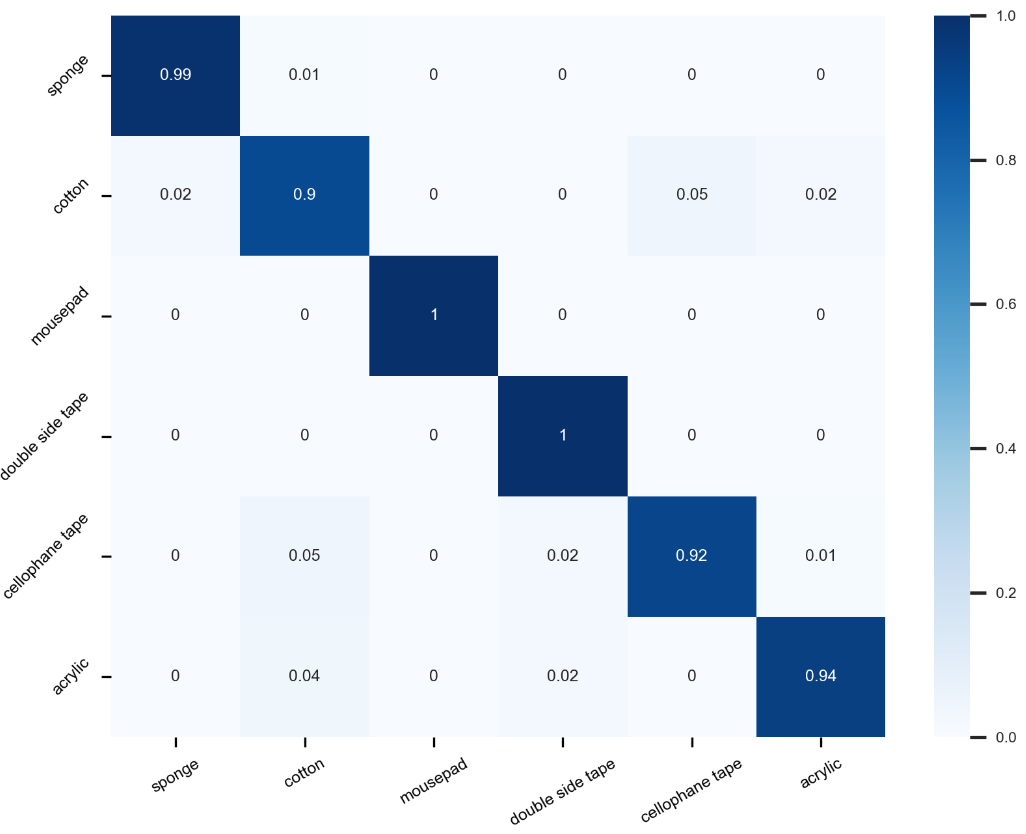}
    \caption{Hardness confusion matrix}
    \label{fig:hardnessConfusionMatrix}
\end{figure}

\subsubsection{Effect of window size}
\begin{figure}[ht]
    \centering
    \includegraphics[width=0.43\textwidth]{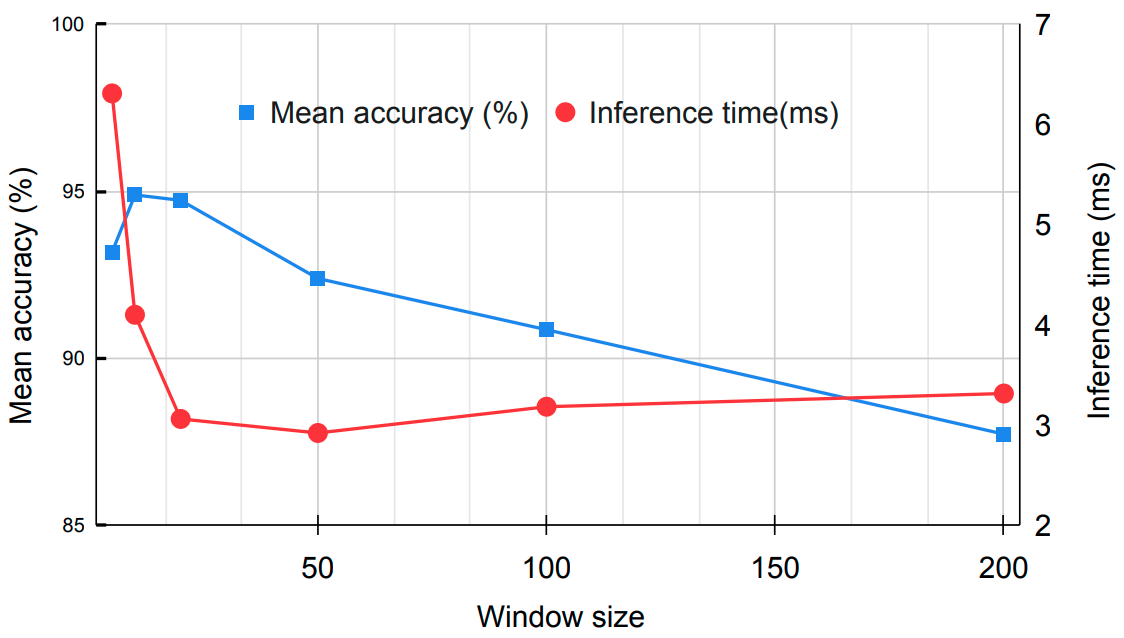}
    \caption{Effect of window size over roughness classification accuracy and inference time.}
    \label{fig:acuracyVsInference}
\end{figure}

\begin{figure}[ht]
    \centering
    \includegraphics[width=0.43\textwidth]{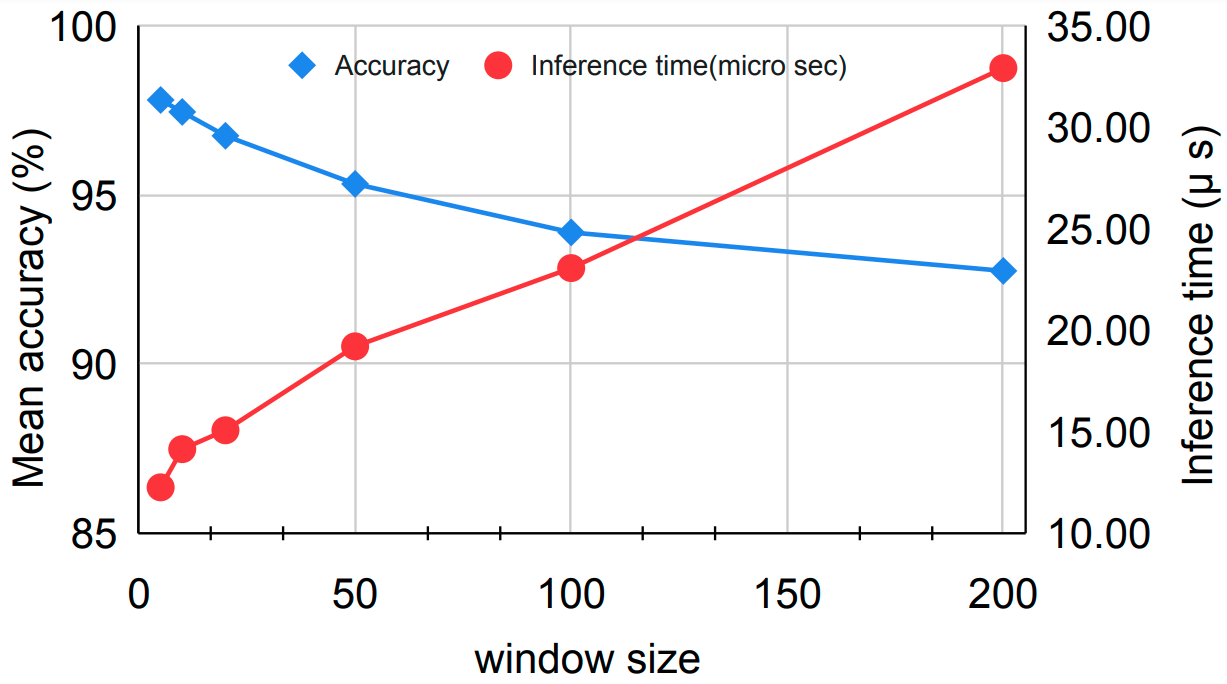}
    \caption{Effect of window size over hardness classification accuracy and inference time}
    \label{fig:hardnessAccuracyInference}
\end{figure}

The temporal window ($W$) introduced in section \ref{sec:notations} plays a critical role in model performance. The length of surface captured in one window is given as $L = \mathcal{D}*W$. Intuitively, a smaller window will have lower dimensional features leading to better machine learning model performance \cite{dim-reduce-better}. However, having a smaller $W$ leads to higher training and inference time due to the increased number of samples for training the model. Therefore, this leads to a trade-off between model performance and inference time, both critical for real-life applications. We evaluate the effect of variation of temporal window size ($W$) on model accuracy, training time, and inference time, as shown in Fig. \ref{fig:acuracyVsInference} and Fig. \ref{fig:hardnessAccuracyInference}. Table \ref{tab:resultsRoughness} \& \ref{tab:resultsHardness} shows the effect of window size.


\subsubsection{Validation with Laser Data}
A comparison between textural roughness classification accuracy of models trained on whisker and laser sensor data reveals equivalent performance. We list configuration for best classification accuracy below:

\begin{itemize}
    \item Whisker Sensor (PA) $\longrightarrow$ $W \:= \:50$, $\mathcal{V}_s \:= \: 50$ : Reported Accuracy (SVM classifier) - $92.60 \%$
    
    \item Laser Sensor $\longrightarrow$ $W \:= \:100$, $\mathcal{V}_s \:= \: 50$ : Reported Accuracy (MLP classifier) - $92.85 \%$
    
\end{itemize}

Machine learning model performance on data collected from the whisker sensor is at par with that of laser. We also see that the PA combination performs best in all cases compared to P and A. These observations are summarized in the following expression.\\[-1em]

\begin{equation}
\label{eq:score-comparison}
(PA_{accuracy} \approx (L)_{accuracy}) > P_{accuracy} > A_{accuracy}
\end{equation}

\subsubsection{Hardness Classification}
Table \ref{tab:resultsHardness} shows the results for texture hardness classification, with excellent classification across all categories as shown in Fig. \ref{fig:hardnessConfusionMatrix}. Previously, we saw that combining accelerometer and pressure sensor data was helpful for texture roughness classification. However, we see that for texture hardness classification, accelerometer data does not aid in improved model performance, leading to performance degradation in some cases. The slip-stick mechanism was responsible for transferring whisker tip information to the base where the accelerometer is located. However, in the dabbing method, the slip-stick mechanism is absent. This leads to very little information about the specimen hardness being transferred to the accelerometer. Therefore, combining accelerometer and pressure sensor data does not lead to increased performance in texture hardness classification.  We list configuration for best classification accuracy below:

\begin{itemize}
    \item Whisker Sensor (PA) $\longrightarrow$ $W \:= \:50$: Reported Accuracy (RF classifier) - $96.42 \%$
\end{itemize}


\begin{table}[ht]
\centering
\caption{Textural Hardness Classification Results}
\scalebox{0.9}{
\begin{tabular}{|*{8}{c|}}  
\hline
\\[-1em]
\multirow{2}{*}{window size} & \multirow{2}{*}{Classifier} &\multicolumn{2}{|c}{\cellcolor{red!10}A}& \multicolumn{2}{|c}{\cellcolor{green!10}P}& \multicolumn{2}{|c|}{\cellcolor{blue!10}PA}\\ \cline{3-8}
\\[-1em]
& &$\mu$ &$\sigma^2$ &$\mu$ &$\sigma^2$ &$\mu$ &$\sigma^2$\\ \hline
\\[-1em]
\multirow{3}{*}{50}& SVM& 41.82& 2.10& 72.59& 2.61& 67.66& 1.18\\ \cline{2-8}
\\[-1em]
                    & RF& 38.48& 2.68& 96.23& 0.39& 96.42& 0.05\\ \cline{2-8}
\\[-1em]
                    & MLP& 26.14& 97.06& 68.02& 4.61& 77.96& 2.35\\ \cline{1-8}
\\[-1em]
\multirow{3}{*}{100}& SVM& 46.10& 3.46& 67.58& 5.04& 70.96& 3.93\\ \cline{2-8}
\\[-1em]
                    & RF& 46.06& 6.05& 95.26& 0.95& 93.77& 0.32\\ \cline{2-8}
\\[-1em]
                    & MLP& 20.28& 5.48& 65.95& 0.50& 73.17& 7.18\\ \cline{1-8}
\end{tabular}
}
\label{tab:resultsHardness}
\end{table}

It is clear from Table  \ref{tab:resultsHardness} that combining accelerometer with pressure sensor data doesn't yield improved model performance. The observations can be summarized in the following expression.\\[-1em]

\begin{equation}
\label{eq:hardness_score-comparison}
    (P_{accuracy}  \approx  (PA)_{accuracy}) > A_{accuracy}
\end{equation}

\subsubsection{Effect of Downsampling}
\begin{figure}[ht]
    \centering
    \includegraphics[width=0.43\textwidth]{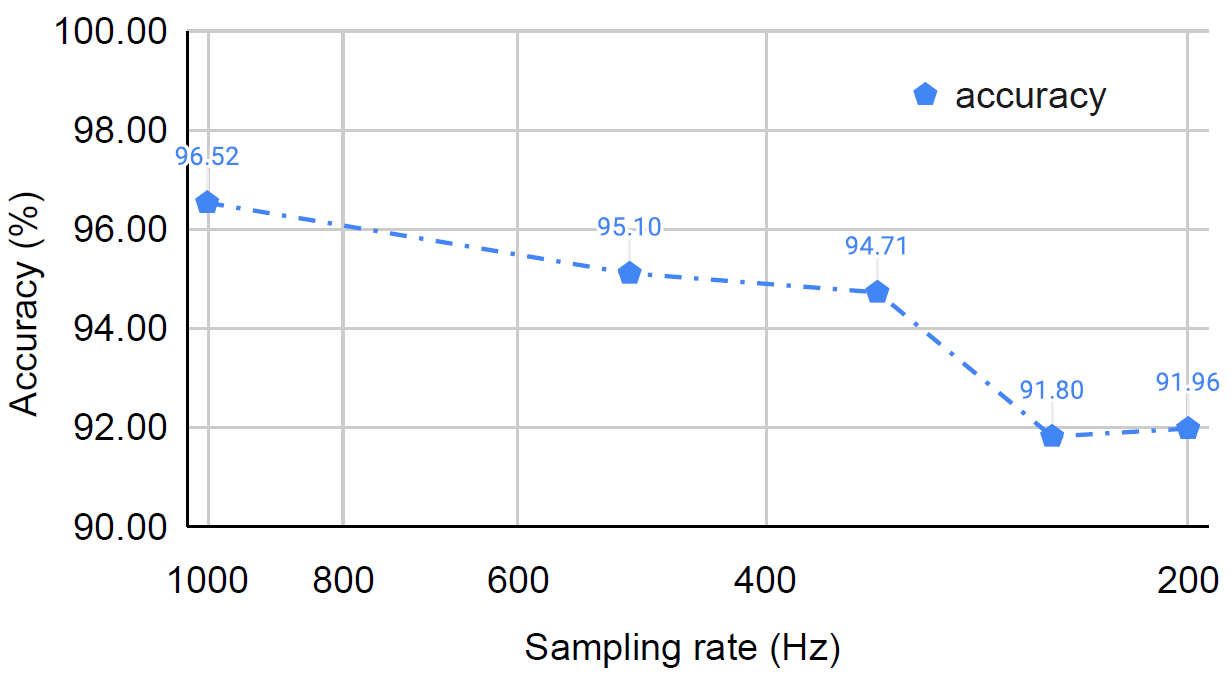}
    \caption{Effect of sampling over the overall accuracy}
    \label{fig:downsampleAccuracy}
\end{figure}

The combination of pressure and accelerometer data used for training the machine learning model was captured at $1000 Hz$. It is important to note that repetition in the pressure sensor data was introduced since  $N_P$ is limited to $157 Hz$. The accelerometer captures information about the texture at a higher temporal resolution. A critical insight is using the accelerometer and the pressure sensor data for texture classification. It allows the machine learning model to discover the missing gaps in texture information leading to better classification performance. To evaluate the premise that a higher texture roughness classification accuracy is due to the higher sampling rate of the accelerometer, we downsample the data to: 1) $1000 Hz$, 2) $500 Hz$, 3) $333 Hz$, 4) $250 Hz$, and 5) $200 Hz$. Fig. \ref{fig:downsampleAccuracy} shows the accuracy for the five different data rates; we have used $\mathcal{V}_{s} = 50 \; and \; W = 50$ based on Table \ref{tab:resultsRoughness}. From Table \ref{tab:resultsRoughness} and Figure \ref{fig:downsampleAccuracy}, we can observe that combining the pressure and accelerometer data at a higher sampling rate allows richer texture information to be captured. It results in higher classification accuracy from around $91.96\%$ to $96.52\%$ for the SVM classifier. A similar trend is observed for other classifiers as well.


\subsubsection{Limitations}
A drawback of all contact-based texture data acquisition systems is that abrupt texture changes can lead to excessive vibrations captured alongside the texture data. This study did not consider external noise introduced in the texture data due to the environment or the textured surface. The added noise can thus degrade the SNR, resulting in a reduction of classification accuracy. A way to circumvent this problem is to have multiple whiskers of varying specifications and to have a voting mechanism between their classification results to determine the correct texture of the specimen. Further, we have considered only symmetric surface texture in our study, and the models are yet to be tested on real terrain. Finally, machine learning models that exploit the texture data's temporal nature should be explored. Previous studies have considered models of whisker-like tactile sensors \cite{tiwari2022visibility}, these abstract models can help design an interface between the raw whisker data and the machine learning model.


\section{Conclusion}
In this study, we introduced a new whisker sensor to capture rich multi-dimensional texture information: roughness and hardness. An experimental setup was designed to capture the surface texture roughness and hardness by sweeping and dabbing methodology. The performance of three machine learning models (SVM, RF, and MLP) showed excellent classification accuracy for surface texture roughness and hardness. Comparing the classification accuracy of three machine learning models trained on whisker data with that of the laser data showed equivalent performance, validating the reliability of the collected whisker sensor data. The results show that the combination of pressure sensor and accelerometer data performs best for textural roughness classification. In contrast, pressure sensor data showed equivalent performance for texture hardness classification. The results also show that accelerometer data contributes to increased classification accuracy due to a higher sampling rate, allowing it to capture surface texture information at a higher temporal resolution.

In future, multiple whiskers of varying specifications and a  voting mechanism could perform better for asymmetric texture surfaces. Finally, machine learning models could be explored to exploit the temporal nature of the texture data.

\bibliographystyle{./bibliography/IEEEtran}
\bibliography{./bibliography/refs}

\end{document}